%% file: manuscript_rev.tex
\documentclass[10pt]{iopart}
\usepackage{iopams}
\usepackage{tikz}
\usepackage{hyperref}  
\usepackage{nicefrac}
\usepackage{upgreek}
\newcommand{\uvec}[1]{\mathbf{\hat{#1}}}
\newcommand{\sech}{\mathrm{\, sech}}
\newcommand{\bm}{\boldsymbol}
\newcommand{\bmat}{\left( \begin{array}{cc}}
\newcommand{\emat}{\end{array} \right)}

\begin{document}


\title[FPUT recurrence in a Heisenberg spin chain]{Exact and non-exact Fermi-Pasta-Ulam-Tsingou recurrences in a Heisenberg ferromagnet}

\author{Rahul O. R. and S. Murugesh}

\address{Department of Physics, Indian Institute of Space Science and Technology, Thiruvananthapuram - 695~547, India.}
\ead{murugesh@iist.ac.in}

\begin{abstract}
We visualize the Fermi-Pasta-Ulam-Tsingou (FPUT) recurrence in a classical Heisenberg ferromagnetic (HF) spin chain by exploiting its gauge equivalence to the nonlinear Schr\"{o}dinger equation (NLSE). We discuss two types of spatially periodic breather excitations in the spin chain, that are associated with: (I) Akhmediev breather, and (II) Galilean transformed Akhmediev breather. The recurrence in the former is exact in the sense that the initial and final states are identical. In the later, the spin chain undergoes an additional global rotation during the recurrence process, which makes the initial and final states distinguishable. Both the complex solutions (I) and (II) nevertheless show a definite phase shift during the recurrence process. A one-to-one correspondence between HF spin chain and the NLSE seems missing by virtue of the closeness of the FPUT recurrence.
\end{abstract}

%
\vspace{2pc}

\submitto{\PS}
%
%
\ioptwocol

\section{Introduction}
A major turning point in the study of nonlinear dynamical systems was the Fermi-Pasta-Ulam-Tsingou (FPUT) numerical experiment performed using one of the earliest computers, \mbox{MANIAC I}, 
in 1955\,\cite{fput:1955}. FPUT experiment has a decisive role in the understanding of integrable systems and soliton dynamics\,\cite{ford:1992}.  
Although the theoretical model under study was quite simple, its results were controversial in nature which continue to inspire many researchers all over the globe even after six decades\,\cite{pier:2018}. 

FPUT numerically studied a coupled system of finite number of one dimensional anharmonic oscillators. The first normal mode was excited initially with a finite amount of energy, expecting the sharing of energy with all the higher modes in equal manner owing to the non-linearity of the system. Sharing does occur as anticipated. However, after a sufficiently long time the initial mode is recovered with nearly the same energy as in the beginning, which can be treated as a near \emph{recurrence}. Later studies showed that the mathematical model considered in the FPUT experiment is in a way an integrable system\,\cite{jacks:1990}. Although nonlinear, it was noted that the motion of an integrable system can repeat, in contrast to a chaotic one. 

One can consider the {recurrence} as the recovery of a smooth (single wave) initial condition through a rather complicated nonlinear evolution. This scenario can be exactly modeled by the \emph{breather} solution of nonlinear Schr\"{o}dinger equation (NLSE) --- more precisely a spatially periodic breather\,\cite{akhm:1986} widely known as Akhmediev breather (AB). In the context of AB, initial and final states are quite steady and smooth. The dynamics is actually taking place in between; much like shuffling playing cards in a peculiar way that the cards end up in the same initial configuration. In the Fourier expansion of AB\,\cite{hammani:2011,devine:2011}, all the higher modes (sidebands) are absent as $t\to \pm\infty$, but they grow and decay during the breather excitation. This resembles the growth--decay cycle of higher normal modes seen in the FPUT recurrence, provided that the recurrence occurs only once\,\cite{akhm:2001}. The NLSE being a model for the nonlinear optical channel, the FPUT recurrence has been experimentally demonstrated in optical fibers\,\cite{vansim:2001}. 

There exist another mathematical model closely related to the NLSE, known as classical Heisenberg Ferromagnetic (HF) spin chain\,\cite{ml:1977}.
In this model, spins can be considered as unit vectors each having 3 rotational degrees of freedom, placed over a one-dimensional lattice. Spins at adjacent sites interact via the exchange interaction arising out of the Pauli exclusion principle\,\cite{van:1932}.
This is an ideal model where there is no external magnetic field or damping force. But the motion is constrained in such a way that nearby vectors try their best to align in the same direction. Interestingly, it is an integrable system endowed with soliton solutions\,\cite{takt:1977}. The relationship between HF model and the NLSE was initially identified by Lakhshmanan\,\cite{ml:1976,ml:1977} through geometrical arguments. More conclusively,  Zakharov and Takhtajan showed that both the systems are gauge equivalent\,\cite{zakh:1979}. They are quite same in their underlying mathematics. Thus all the features of the NLSE can be expected in this continuum spin chain in an entirely different language. What is special about spin chain is that they live in three dimensional space. Unlike complex solutions of the NLSE one can actually ``see'' the dynamics.  

The aim of this paper is to visualize the recurrence process in the HF spin chain. 
This can be achieved through a spatially periodic breather excitation. The spin chain surely goes back to its initial orientation after the breathing event. 
For the NLSE, the recurrence need not be an `exact' recurrence\,\cite{kuznetsov:2017}. For instance, in the context of AB, the initial and final states are not identical; the background field undergoes a definite phase shift which ranges from $0$ to $2\pi$ depending on the spatial periodicity of the breather\,\cite{devine:2011}. In view of this, we examine breather excitation in the HF spin chain to see how close the recurrence is. In other words, we investigate whether or not the initial and final configuration of the spin chain are exactly identical. In this paper, two distinct spatially periodic breathers in the HF spin chain are studied in detail: one corresponds to AB and the other corresponds to Galilean transformed AB. We show that the recurrence in the former is `exact' in the sense that the initial and final configuration are one and the same. In the later, final state of the spin chain is different from the initial one through an additional global rotation. 

\section{Classical HF model and the NLSE}
Dynamics of the Heisenberg ferromagnet is governed by the Hamiltonian\,\cite{van:1945, ashc:1976},
\begin{equation}
\label{hfham}
\mathcal{H} = - J \sum_{i}^{N}  \bi{S}_i \cdot \bi{S}_{i+1} 
\end{equation}
where $\bi{S}_i$ is the quantum mechanical operator for the spin at the $i$-th lattice site and $J$ is the exchange integral. 
$\mathcal{H}$ is invariant under global rotations in the spin space as the energy of interaction depends only on the relative orientation of the nearest neighbours. This is widely known as Heisenberg Hamiltonian or Heisenberg model. $J>0$ leads to the spin ordering in ferromagnets and $J<0$ corresponds to antiferromagnets. One can examine the classical limit of \eref{hfham}, say classical HF model, 
wherein the distance between nearest neighbours approaches zero. Then $\bi{S}_i $ can be replaced by a continuous function $\uvec{S}(\bi{r},t)$, where $\uvec{S} = (S_1 , S_2 , S_3)$ is treated as unit $3$-vector, after rescaling $J$ appropriately. The dynamics of the system is described by Landau Lifshitz Equation (LLE)\,\cite{ml:1976,land:1935,herr:1951}, written as 
\begin{equation}
\frac{d}{dt} \uvec{S}(\bi{r},t) = \uvec{S}(\bi{r},t) \times \nabla^2 \uvec{S}(\bi{r},t) ; \qquad |\uvec{S}|^2=1.
\end{equation} 
In this paper we will be interested in the (1+1) - dimensional case ($x$ and $t$) of the classical HF model, 
\begin{equation}
\label{hfe}
\uvec{S}_t = \uvec{S} \times \uvec{S}_{xx} ; \qquad |\uvec{S}|^2=1,
\end{equation} 
owing to its integrability\,\cite{takt:1977} through Inverse Scattering Transform (IST). Suffix $x$ and $t$ implies partial derivative. Indicating the applicability of IST, a nonlinear equation can be expressed via a system of linear equations known as Lax pair\,\cite{jacks:1990}. Lax pair for the HF spin chain (or 1-d LLE) is given by\,\cite{zakh:1979}
\begin{equation}
\label{laxp_lle}
\eqalign{
\bm{\Phi}_x &= U_{HF} \bm{\Phi} = i\lambda{\bf S}\ \bm{\Phi} ,\\
\bm{\Phi}_t &= V_{HF} \bm{\Phi} = \big{(}\lambda{\bf S}{\bf S}_x +2i\lambda^2{\bf S}\big{)} \bm{\Phi},
}
\end{equation}
where $ {\bf S} = \sum S_j\sigma_j, $ with $\sigma_j,\,\,j=1,2,3,$ being the Pauli matrices and $S_j$ the components of the unit spin field $\uvec{S}(x,t)$. The constant $\lambda$ is the scattering parameter in the language of IST. The compatibility condition $\bm{\Phi}_{xt} = \bm{\Phi}_{tx}$, of the system  \eref{laxp_lle} can be shown to be
\begin{equation}
\label{hfe_matrix}
{\bf S}_t = \frac{1}{2i}[{\bf S}, {\bf S}_{xx}] ; \qquad \mathbf{S}^2 = \mathbf{I},
\end{equation}
which is the matrix form of classical HF model given in \eref{hfe}. In the same manner, the one dimensional focussing NLSE,
\begin{equation}
\label{nlse}
i\psi_{t} + \psi_{xx} + 2 {|\psi|}^2 \psi = 0 ,
\end{equation}
is integrable through IST\,\cite{zakh:1972}, and can be written as the compatibility condition for the linear system
\begin{equation}
\label{laxp_nlse}
\eqalign{
\bm{\Psi}_x &= U \bm{\Psi} ,\\
\bm{\Psi}_t &= V \bm{\Psi} ,
}
\end{equation}
for suitable connection coefficients $U$ and $V$. Although $U$ and $V$ are in general, functions of $x, t$ and $\lambda$, they can be conveniently written as a function of $\psi$ and its derivatives (see  \ref{app:DT} for details). Thus, one can associate a complex solution $\psi$ to the corresponding matrix solution $\bm{\Psi}$, $(\psi_0, \bm{\Psi}_0) $, $(\psi_1, \bm{\Psi}_1) $ etc.  The spin configuration obtained using the relation\,\cite{zakh:1979}
\begin{equation}
\label{gauge}
{\bf S} = \lim_{\lambda\to 0}\bm{\Psi}^{\dagger}\sigma_3\bm{\Psi},
\end{equation}
satisfies \eref{hfe_matrix}, and hence \eref{hfe}. The NLSE and classical HF spin chain are thus said to be gauge equivalent. In short, \eref{gauge} presents a systematic way of finding a spin configuration corresponding to each solution of the NLSE. 
Energy density of the spin field can be expressed as\,\cite{zakh:1979},
\begin{equation}
\label{endens}
E = \uvec{S}_x^2 = 4 |\psi|^2 .
\end{equation}

\begin{figure}[]
\begin{tikzpicture}
\node[inner sep=0] (secant) at (0,0) {
\includegraphics[trim = 0cm 1.8cm 0cm 1.4cm, clip, width=1.0\linewidth]{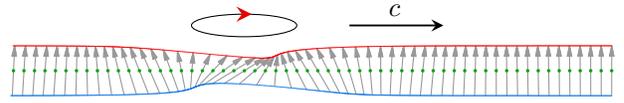}};
\draw[thin] (-0.9,0.6) ellipse (0.7cm and 0.15cm);
\draw[line width=1.5pt,red!90!black,-stealth](-0.9,0.75)--(-0.8,0.75);
\draw[line width=0.7pt, black](0.5,0.6)--(1.7,0.6);
\node[] at (1.1,0.8) {$c$};
\draw[line width=1pt, black,-stealth](1.7,0.6)--(1.75,0.6);
\definecolor{aqua}{rgb}{0.0, 1.0, 1.0}
\end{tikzpicture}
\caption{Soliton in the classical HF spin chain.}
\label{fig:hf_sech}
\end{figure}
The trivial solution $\psi_0 = 0 $ has its counterpart $\uvec{S}_0 = 0 \uvec{i} + 0 \uvec{j} + 1 \uvec{k}$, which is a static field. Energy density is zero because the spins are strictly parallel throughout the spin chain. Starting from this zero \textit{seed} solution, a 1-soliton can be obtained through IST or any other technique for obtaining soliton solutions. We note that, Darbox transformation technique\,\cite{schief:2002,chao:2005} is quite useful for this study, since it generates matrix function $\bm{\Psi}$ along with the soliton solution, and can be iterated as follows:
$
(\psi_0, \bm{\Psi}_0) \longrightarrow (\psi_1, \bm{\Psi}_1) \longrightarrow \cdots
$

The 1-soliton of \eref{nlse} is the well known travelling wave of `secant-hyperbolic' form\,\cite{zakh:1972}
\begin{equation}
\label{sechyp}
\psi_{\rm sh} = 2\lambda_{0I} \exp{(i \mu_{0R})} \sech\, \mu_{0I},
\end{equation}
where $\mu_0 = \mu_{0R} + i \mu_{0I} = 2\lambda_0 x - 4 \lambda_0^2 t$, and $\lambda_0 = \lambda_{0R} + i \lambda_{0I}$ is the complex scattering parameter that determines the amplitude and velocity of the soliton. 
For a qualitative description of soliton in the HF spin chain, notice the $z$-component of the corresponding spin field\,\cite{tjon:1977,takt:1977,murug:2005}
\begin{equation}
S_3 = 1 - \frac{2\lambda_{0I}^2}{|\lambda_0|^2} \sech^2(2\lambda_{0I} x - 8 \lambda_{0I}\lambda_{0R} t),
\end{equation}
which tends to `$1$' except for a small region of the spin chain at any instant of time. Hereafter, $x$ is measured along the lattice axis over which spin vectors are being fixed. This localized disturbance (or excitation) in a 1-d isotropic spin field travels with a uniform speed as shown in \fref{fig:hf_sech}. Energy density $ E \propto \sech^2(x-ct)$, where $c$ is the speed\,\cite{ml:1977}, and $\int\!E\ {d\!x} $ is finite even for an infinite spin chain. 

\section{Seed solutions, breathers and their Galilean transformations}
Our study deals with two kinds of breather solutions. For a better comparison of the results we will classify the complex solutions by invoking their geometrical connection. One can think of a space curve parametrized by its arclength $x$, for which the unit tangents are the spin vectors $\uvec{S}(x)$ satisfying \eref{hfe}. The intrinsic equation describing this moving space curve can be expressed as a coupled partial differential equation of its curvature and torsion, given by\,\cite{ml:1976}
\begin{equation}
\label{darios}
\eqalign{
		\kappa_t &= -2 \kappa_x \tau - \kappa \tau_x  , \\
		\tau_t   &= \left( {\kappa_{xx}}/{\kappa} - \tau^2  \right)_x + \kappa \kappa_x .
		}
\end{equation}
This is Da Rios equation\,\cite{ricc:1996} derived much earlier in an entirely different context.  The parallelism between Da Rios equation and the NLSE can be understood using the Hasimoto function\,\cite{hasi:1972,ml:1977}
\begin{equation}
\label{hasi}
\psi = \frac{1}{2} \kappa \, e^{i \sigma} , \quad \sigma_x = \tau ,
\end{equation}
where $\kappa$ and $\tau$ are respectively, the curvature and torsion of a curve. The complex function $\psi$ is described by the NLSE given in \eref{nlse}, provided that $\kappa$ and $\tau$ obeys the Da Rios equation given in \eref{darios}. Using \eref{hasi} we classify seed solutions as shown in \tref{tbl}.

\begin{table}[]
\caption{Geometric classification of some seed solutions.}\label{tbl}
\begin{indented}
\item[]\begin{tabular}{@{}cccc}
\br
$\psi$ & $\kappa$ 	& $\tau $ 	& Curve \\
\mr
$0$  & $ 0 $ 	& $ 0 $		& line		\\
$ \kappa_0 e^{2 i \kappa_0^2 t } $ 	& $2\kappa_0$& $ 0 $	& circle 	\\
$ \kappa_0 e^{i \sqrt{2}\kappa_0 x }$&$2\kappa_0$& $ \sqrt{2} \kappa_0 $ 	&  helix \\
\br
\end{tabular}
\end{indented}
\end{table}

Apart from solitons, the NLSE also allows \textit{breather} solutions. A solution is said to be a breather if the field variable is localized, and in addition has a periodic nature; either in space or in time. There exist some well known breather solutions to the NLSE: the time periodic Kuznetsov-Ma breather(KMB)\,\cite{kuzn:1977,ma:1979}, spatially periodic Akhmediev breather (AB)\,\cite{akhm:1986} and a special case of both --- Peregrine soliton (PS)\,\cite{pere:1983}, for which periodicity in space and time  tend to infinity. The stability analysis of various breathers has been a topic of intense research, see for instance \,\cite{alejo:2019} and references therein.
A breather solution is obtained using Darboux transformation (or any other standard method for obtaining soliton solutions), if one start with a seed solution
\begin{equation}
\label{seedcircle}
\psi_{\rm c} = \kappa_0\, e^{2i \kappa_0^2 t},
\end{equation}
for an arbitrary real constant $\kappa_0$. Suffix $\rm{c}$ indicates `circle' (refer \tref{tbl}). Seed solution $\psi_{\rm c}$ acts as a  uniform background for the breather solution, in contrast to the zero background as in the case of `secant-hyperbolic' soliton. We will focus only on the spatially periodic breather, i.e.,  Akhmediev breather which may be written as\,\cite{dysthe:1999}
\begin{eqnarray}
\label{psiab}
\psi_{\rm AB}(x,t) = \kappa_0 e^{2 i \kappa_0^2 t} \nonumber\\ \qquad \qquad \quad
\times \ \frac{\cosh(rt-2i\phi) - \cos\phi \ \cos(qx)}{\cosh(rt) - \cos\phi \ \cos(qx)},
\end{eqnarray}
where, $q = 2 \kappa_0 \sin\phi$, $r= 2 \kappa_0^2 \sin(2\phi)$ and $\phi \in [0,\pi/2]$ is a real parameter. The spatial periodicity of \eref{psiab} is $2\pi / q$, which tends to infinity as $\phi \to 0$; the rogue behaviour. Breather becomes the seed solution as $\phi \to \pi/2$.

The breather solution over the background
\begin{equation}
\label{seedhelix}
\psi_{\rm h}  =\kappa_0\, e^{i\sqrt{2}\, \kappa_0 x},
\end{equation}
for an arbitrary real constant $\kappa_0$, has been recently obtained by the authors\,\cite{rahulhelix}, and shown to have a \emph{knotted} structure in the associated space curve. Spatially periodic case of this new breather is a Galilean transformed version of \eref{psiab}, which may be written as
\begin{eqnarray}
\label{psigab}
\psi_{\rm GAB}(x,t) = \kappa_0\, e^{i \sqrt{2} \kappa_0(x-x_0)} \nonumber\\ \qquad \quad
\times \ \frac{\cosh(rt-2i\phi) - \cos\phi \ \cos(q(x-vt))}{\cosh(rt) - \cos\phi \ \cos(q(x-vt))} ,
\end{eqnarray}
where $v=2\sqrt{2} \kappa_0 $, $x_0 = \pi/q$, and $q, r, \phi$ are defined below \eref{psiab}. Refer  \ref{app:GAB} for details. Galilean transformed Akhmediev Breather --- $\psi_{\rm GAB}$, were obtained earlier by Salman\,\cite{salman:2013}. These are a special case of the more general breather solution over the plane wave $\psi_{\rm h}$, upto a scaling and suitable choice of parameters.  

It may be noted that under the Galilean transformation
\begin{equation}
\label{galilean0}
x\to x-vt, \quad t\to t,
\end{equation}
the NLSE is invariant\,\cite{dysthe:1999} through an additional phase change to the field function,
\begin{equation}
\label{galilean1}
\psi \to \psi \exp{[i( vx/2 - v^2t/4 + v_0 )]} ,
\end{equation}
where $v$ and $v_0$ are arbitrary real constants. One can verify that the plane wave solution $\psi_{\rm c}$ transforms to $\psi_{\rm h}$ under the Galilean transformation for $v = 2\sqrt{2}\kappa_0$ and $v_0 = 0$. In terms of associated space curve (refer \tref{tbl}), the circle transforms to a helix. Specifically, the curve picks up an additional torsion under the Galilean transformation.

We will discuss the spin field associated with $\psi_{\rm AB}$ and $\psi_{\rm GAB}$ in the coming sections. 

\section{Exact recurrence in classical HF spin chain}
\label{sec:ab}
The spin configuration associated with $\psi_{\rm c}$ can be written in the vector form as\,\cite{rahulhf}
\begin{equation}
\label{Sc}
\uvec{S}_{\rm c} =\cos(2\kappa_0x) \uvec{j} + \sin(2\kappa_0x) \uvec{k}.
\end{equation}
Two constants indicating global rotation and translation are set to zero  without loss of generality. This ``seed'' spin is a static field (independent of $t$) with energy density $4\kappa_0^2$. Recently, Darboux transformation 
has been constructed starting with a more general seed solution\,\cite{zhang:2014,zai:2016} for which  \eref{Sc} arises as a special case. 
A comparative study between classical HF model and the NLSE, in the context of breather (rogue) excitations has also been done recently\,\cite{aritra:2015}, wherein the space curve formalism has been employed to obtain the curvature and torsion associated with NLSE breather solutions.
Apart from the simplest form discussed in this paper, the classical HF model with additional interactions has been investigated recently using Darboux transformation, in connection with the breathers of higher order NLSE\,\cite{jian:2018, yang:2017}.
In view of the above developments, our study goes beyond this by exploring the recurrence phenomena in the spin chain.

A general expression for spin breather over the background $\uvec{S}_{\rm c}$, for the spectral parameter $\lambda_0 = \lambda_{0R} + i \lambda_{0I}$ is obtained in \,\cite{rahulhf}. The spatial periodicity $L_1=\pi/\kappa_0$ of $\uvec{S}_{\rm c}$ must appear in the spin breather. We consider here, the spatially periodic breather, say $\uvec{S}_{\rm AB}$ for which a periodicity of $L_2 = 2\pi / q$ also arises as in the case of $\psi_{\rm AB}$ given in \eref{psiab}. It can be seen that $L_1/L_2 = \sin \phi$. The parameter $\phi$ that determines the spatial periodicity in \eref{psiab} can be shown to have a dependence on $\kappa_0$ and $\lambda_{0I}$. In terms of $\lambda_0$, the condition for AB may be written as $\lambda_{0R} = 0$ and $\kappa_0 > \lambda_{0I}$. It follows from \,\cite{rahulhf} that $L_1/L_2 = \sqrt{1-(\lambda_{0I}/\kappa_0)^2}$. This readily gives 
\begin{equation}
\label{cosphi}
\cos \phi  = \lambda_{0I}/\kappa_0, 
\end{equation}
and the rogue phenomena then corresponds to a case where $\lambda_{0I} \sim \kappa_0$. 

\begin{figure}[]
\centering
\begin{tikzpicture}
\node[inner sep=0] (spingab) at (0,0) {
\includegraphics[trim = 0.2cm 6.8cm 0.2cm 5.0cm, clip, width=0.92\linewidth]{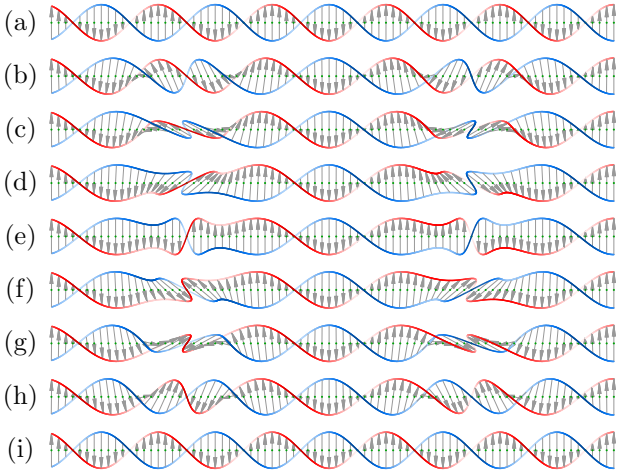}};
\def\x{-4.15} 
\def\y{2.7} 
\def\h{0.71} 
\node[] at (\x,\y) {(a)}; \pgfmathsetmacro\y{\y - \h};
\node[] at (\x,\y) {(b)}; \pgfmathsetmacro\y{\y - \h};
\node[] at (\x,\y) {(c)}; \pgfmathsetmacro\y{\y - \h};
\node[] at (\x,\y) {(d)}; \pgfmathsetmacro\y{\y - \h};
\node[] at (\x,\y) {(e)}; \pgfmathsetmacro\y{\y - \h};
\node[] at (\x,\y) {(f)}; \pgfmathsetmacro\y{\y - \h};
\node[] at (\x,\y) {(g)}; \pgfmathsetmacro\y{\y - \h};
\node[] at (\x,\y) {(h)}; \pgfmathsetmacro\y{\y - \h};
\node[] at (\x,\y) {(i)};
\end{tikzpicture}
\caption{Breather mode in classical HF spin chain in its time evolution\,\cite{rahulhf}. This is associated with AB, \eref{psiab}. Evolution of the spin chain begins with a uniform background, then gradually develops localized excitations that are spaced periodically along the lattice, and eventually vanishes to get back the same uniform initial background as in the beginning. The initial and final states
(as $t\rightarrow \pm\infty$) seen in (a) and (i) are indistinguishable. Here, $\kappa_0 = 1$, and length of the chain $L = 5 L_1 = 2 L_2$.
}
\label{fig:spin_ab}
\end{figure}
The recurrence can be studied in a spin chain of arbitrary length.  However, for a better description we assume a condition, $\uvec{S}_{\rm AB}(x+L,t)=\uvec{S}_{\rm AB}(x,t)$, in such a way that 
\begin{equation}
L = n L_1 = m L_2,
\end{equation}
where $m,n \in \mathbb{Z}^+$, and $n>m>0$. 
Time evolution of the spin breather for $n=5$ and $m=2$ is presented in \fref{fig:spin_ab}. The background spin field seen in (a) and (i) has $n$ repeating segments. Over this background, the localized excitations can be seen at $m$ places during the breathing event. What is narrated in the time evolution is the recurrence process. It is clear from \fref{fig:spin_ab} that the initial and final states of the spin chain are identical. No trace of the excitation is left in the system after the breathing event. Hence this is an exact recurrence.

\begin{figure}[]
	\centering
	\input{psi_ab.tex}
	\caption{Breather profile of AB, 
\eref{psiab}. For comparison, the parameter values are same as that of the corresponding spin breather shown in \fref{fig:spin_ab}. 
The energy density of the spin chain $E = \uvec{S}_x^2 = 4 |\psi|^2$ has the same profile upto a scaling.}
\label{fig:psi_ab}
\end{figure}
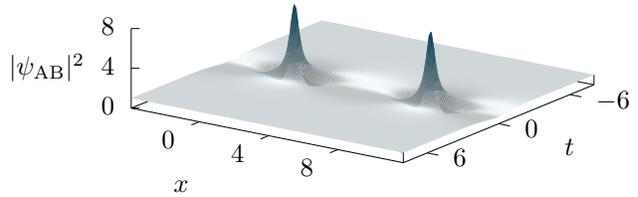

One may compare the localized excitations seen in the spin chain with the corresponding breather profile of $\psi_{\rm AB}$ shown in \fref{fig:psi_ab}. It is obvious that, as $t\rightarrow \pm\infty$, the magnitute, $|\psi_{\rm AB}|$, approaches the constant value $\kappa_0$. However, the phase $ \varphi = \rm{arg}(\psi_{\rm AB})$, shows a definite shift when comparing at $t\rightarrow -\infty$ and $t\rightarrow +\infty$. This has been studied in detail by Devine \etal\,\cite{devine:2011} and shown that the background field ($e^{it}$) undergoes a phase shift, $\Delta \varphi\in [0, 2\pi]$, depending on the spatial periodicity of the breather. It may be noted that the time dependence of the background field ($e^{it}$) is omitted from the phase analysis, as to emphasize the contribution from breather excitation. In the limiting case, Peregrine soliton is marked by a phase shift of $2\pi$. Recently, phase evolution under Peregrine soliton has been experimentally observed in water wave and optical fibers\,\cite{gang:2019}. 

It is worth mentioning that the spin chain in \fref{fig:spin_ab}~(a) can be thought of as a belt having $n$ turns between its ends. The continuous evolution leads to a configuration seen in \fref{fig:spin_ab}~(e) wherein the net turns become $n-2m$. In a finite spin chain, the rogue event is for $ n=2$ and $m=1 $, which then shows a transition from `two' net turns to `zero' net turn\,\cite{rahulhf}. This is a manifestation of the well known Dirac belt trick\,\cite{staley:2010} that demonstrates the triviality of $4\pi$ rotation.

In the following  section we focus on the breather solution over the plane wave $\psi_{\rm h}$, \eref{seedhelix}. The corresponding spin breather in the analytical form is presented and its recurrence process is discussed. We will show that the recurrence is not `exact' in contrast to the one shown in this section.

\section{Spin breather over a rotating spin chain}
\label{sec:spinbr}
Spin configuration associated with $\psi_{\rm h}$ can be obtained by a lengthy, yet straight forward calculation as detailed in \ref{app:DT}. We write it explicitly in the vector form as
\begin{equation}
\label{Sh}
\uvec{S}_{\rm h} = \sqrt{\nicefrac{2}{3}} \, \Big(
			(\nicefrac{1}{\sqrt{2}\,}) 							\, \uvec{i}
			+ \cos \theta 									\, \uvec{j}
			+ \sin \theta 									\, \uvec{k} \Big) ,
\end{equation}
where $\theta = \sqrt{6}\kappa_0(x-\sqrt{2}\kappa_0 t)$. This spin configuration has a spatial period $P_1 = \sqrt{2}\pi/(\sqrt{3}\kappa_0)$ and a uniform rotation about the $x$ -- axis (lattice axis) with time period $T_0=\pi/(\sqrt{3}\kappa_0^2)$. Hence this spin field is a dynamical field. The energy density $ 4\kappa_0^2$ is uniform throughout the chain. 

Analytical expression for the breather solution $\psi_{\rm hb}$ starting from $\psi_{\rm h}$  were obtained recently by the authors in \,\cite{rahulhelix}. Suffix `$\rm{hb}$' may be read as breather over the helix. Here we extend the result towards classical HF spin chain using the gauge equivalence, given in \eref{gauge}. Details of the calculation is given in \ref{app:DT}.
Explicit expression for the spin breather is obtained as
\begin{eqnarray}
\label{spin_hb}
\uvec{S}_{\rm hb}(x,t) = \frac{\lambda_{0R}^2-\lambda_{0I}^2}{|\lambda_0|^2} \uvec{S}_{\rm h}
\nonumber \\ 
+ \Bigg{[} - \frac{2 \lambda_{0I}^2 \xi }{|\lambda_0|^2 \chi^2}
					\bigg(- \frac{(\sqrt{2} \eta + \xi )}{\sqrt{3}} 	\bigg)
+ \frac{2 \lambda_{0I} \lambda_{0R}}{|\lambda_0|^2 \chi} 
				\bigg( -\frac{\sqrt{2}}{\sqrt{3}}  \zeta 			\bigg)
\Bigg{]}\uvec{i} \, 
\nonumber \\
+ \Bigg{[} - \frac{2 \lambda_{0I}^2 \xi}{|\lambda_0|^2 \chi^2} 
				\bigg(\cos \theta \frac{(\eta - \sqrt{2} \xi)}{\sqrt{3}} - \zeta \sin \theta \bigg) 
\nonumber \\ \qquad \qquad 
+ \frac{2 \lambda_{0I} \lambda_{0R}}{|\lambda_0|^2 \chi} 
				\bigg(  \frac{\zeta}{\sqrt{3}} \cos \theta + \eta \sin \theta \bigg) 
\Bigg{]}\uvec{j} \quad 
\nonumber \\
+ \Bigg{[} - \frac{2 \lambda_{0I}^2 \xi}{|\lambda_0|^2 \chi^2}
				\bigg(\sin \theta \frac{(\eta - \sqrt{2} \xi)}{\sqrt{3}} + \zeta \cos \theta \bigg)
\nonumber \\ \qquad \qquad
+ \frac{2 \lambda_{0I} \lambda_{0R}}{|\lambda_0|^2 \chi}
				\bigg(\frac{\zeta}{\sqrt{3}} \sin \theta  - \eta \cos \theta \bigg)
\Bigg{]}\uvec{k}, \ \ 
\end{eqnarray}
wherein,\\
$\zeta 	= c_1 \, \cos(\Omega_{0R}) + c_2 \, \cosh(\Omega_{0I}), $ \\
$\eta 	= c_3 \, \sin(\Omega_{0R}) - c_4 \, \sinh(\Omega_{0I}), $ \\
$\xi 	= c_4 \, \sin(\Omega_{0R}) + c_3 \, \sinh(\Omega_{0I}), $ \\
$\chi 	= c_2 \, \cos(\Omega_{0R}) + c_1 \, \cosh(\Omega_{0I}), $ \\
$\Omega_0 = \Omega_{0R} + i\, \Omega_{0I} = 2\, f_0 \, ( x - \sqrt{2} \mu_0 t ), $\\
$f_0 	= f_{0R} + i f_{0I} =  (\nicefrac{1}{\sqrt{2}}) \, \sqrt{\nu_0^2 + 2\, \kappa_0^2}, $ \\
$\mu_0 	= \mu_{0R} + i\mu_{0I} =  \kappa_0 - \sqrt{2}\, \lambda_0, $ \\$
\nu_0 	= \nu_{0R} + i\nu_{0I} =  \kappa_0 + \sqrt{2}\, \lambda_0,$ \\
$c_1 = 2 \, \big( 4\, \kappa_0^2 + 2\, |\nu_0|^2 + 4 \sqrt{2}\, \kappa_0 \, \nu_{0I} + 4\, |f_0|^2 \big),
\, $\\
$c_2 = 2 \, \big( 4\, \kappa_0^2 + 2\, |\nu_0|^2 + 4 \sqrt{2}\, \kappa_0 \, \nu_{0I} - 4\, |f_0|^2 \big),
 $\\
$c_3 = 2 \, \big( 8\, \kappa_0 \, f_{0I} + 4\sqrt{2}\, (\nu_{0R} \, f_{0R} + \nu_{0I} \, f_{0I})\,\big),
$\\
$c_4 = - 2 \, \big( 8\, \kappa_0 \, f_{0R} + 4\sqrt{2}\, (\nu_{0I} \, f_{0R} - \nu_{0R} \, f_{0I})\,\big),$ \\
and $\lambda_0 =\lambda_{0R}+i\lambda_{0I}$ is the scattering parameter in the framework of IST. The background spin field $\uvec{S}_{\rm h}$ (and $\theta$) is defined in \eref{Sh}. The above functions and constants also obey the conditions:
\begin{equation}
\label{cons}
\zeta^2 + \eta^2 + \xi^2 = \chi^2, \mbox{and} \quad  
c_2^2 +c_3^2+c_4^2 =  c_1^2. 
\end{equation}

The corresponding breather solution to the NLSE may be written as\,\cite{rahulhelix}
\begin{equation}
\label{psi_hb}
\psi_{\rm hb}=  e^{i \sqrt{2} \kappa_0 x} \Big(\kappa_0 - 2\, \lambda_{0I}\, {(\zeta - i\, \eta )}\big/{\chi} \, \Big),
\end{equation}
where the functions and parameters are same as that of \eref{spin_hb} (See \ref{app:DT} for details). This is a travelling breather which is periodic both in space and time. One can infer that the localization arises from hyperbolic functions in $\zeta, \eta, \xi$ and $\chi$ (given below \eref{spin_hb}) for which the argument is $\Omega_{0I}$. On the other hand, periodicity comes from $\Omega_{0R}$ via trigonometric functions. Of particular interest here, we focus on a special case where $\Omega_{0I} \equiv \Omega_{0I}(t)$, a function of $t$ alone, so that the breather peaks align at $t=0$ line in the $x-t$ plane. This spatially periodic breather can be obtained by the condition
\begin{equation}
\label{cond_GAB}
\lambda_{0R} = - {\kappa_0}\big/{\sqrt{2}} \ , \ \mbox{and}\quad  \kappa_0^2 > \lambda_{0I}^2.
\end{equation}
With this condition the general breather given in \eref{psi_hb} reduces to $\psi_{\rm GAB}$ given in \eref{psigab}. Detailed steps are provided in \ref{app:GAB}. One can see that the above condition leads to \eref{cosphi} as in the case of AB, because both $\psi_{\rm AB}$ and $\psi_{\rm GAB}$ have the same spatial periodicity.
\section{Recurrence with a global rotation}
\label{sec:gab}
\begin{figure}[]
\centering
\begin{tikzpicture}
\node[inner sep=0] (spingab) at (0,0) {
\includegraphics[trim = 0.5cm 6.8cm 0.2cm 5.7cm, clip, width=0.95\linewidth]{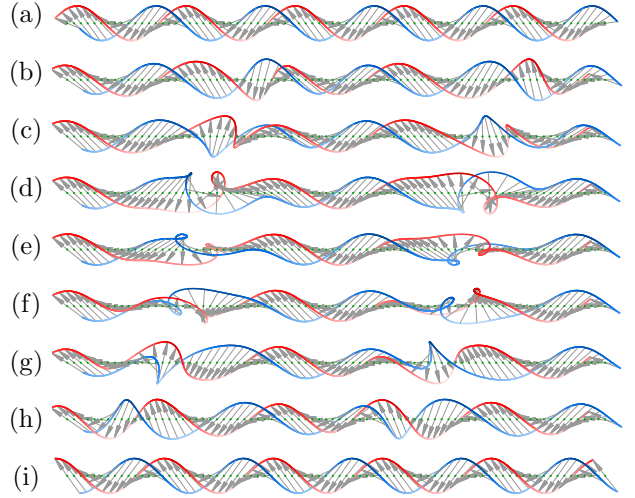}};
\def\x{-4.3} 
\def\y{3.2} 
\def\h{0.77} 
\node[] at (\x,\y) {(a)}; \pgfmathsetmacro\y{\y - \h};
\node[] at (\x,\y) {(b)}; \pgfmathsetmacro\y{\y - \h};
\node[] at (\x,\y) {(c)}; \pgfmathsetmacro\y{\y - \h};
\node[] at (\x,\y) {(d)}; \pgfmathsetmacro\y{\y - \h};
\node[] at (\x,\y) {(e)}; \pgfmathsetmacro\y{\y - \h};
\node[] at (\x,\y) {(f)}; \pgfmathsetmacro\y{\y - \h};
\node[] at (\x,\y) {(g)}; \pgfmathsetmacro\y{\y - \h};
\node[] at (\x,\y) {(h)}; \pgfmathsetmacro\y{\y - \h};
\node[] at (\x,\y) {(i)};
\end{tikzpicture}
\caption{Another breather mode in classical HF spin chain, which is a spatially periodic case of  \eref{spin_hb} obtained 
via \eref{cond_GAB}. This is associated with the Galilean transformed  AB, \eref{psigab}. Uniform rotation (in time) present in the background spin $\uvec{S}_{\rm h}$, \eref{Sh}, has been cancelled out for simplicity. Evolution of the spin chain begins from a uniform background, gradually develops localized excitations and eventually vanishes to get back the same uniform initial background. Initial and final state, (a) and (i) are not identical; the spin chain undergoes an additional global rotation about the lattice axis during the breather excitation. The recurrence is not `exact' in contrast to the similar recurrence shown in \fref{fig:spin_ab}. Here, $\kappa_0 = 1$, and the length of the chain $P = 5 P_1 = 2 P_2$.
}
\label{fig:spin_gab}
\end{figure}
Spatially periodic spin breather associated with $\psi_{\rm GAB}$ has two distinct  periods given by
\begin{equation}
\label{helix_p12}
P_1 = {\sqrt{2}\pi}\big/{(\sqrt{3}\kappa_0)} , \ \ P_2 = {2\pi}/{q} = {\pi}\Big/{\sqrt{\kappa_0^2 - \lambda_{0I}^2}}. 
\end{equation}
The former comes from \eref{Sh} and later comes from \eref{psigab}. In order to have a period matching in a finite spin chain of length $P$, we choose two integers $n, m \in \mathbb{Z}^+$ such that,
\begin{equation}
\label{helix_P}
P = nP_1 = mP_2.
\end{equation}
For a given value of $ \kappa_0$, parameter $\lambda_{0I}$ may be written as
\begin{equation}
\lambda_{0I} = \kappa_0 \sqrt{1- ({3m^2}/{2n^2})},
\end{equation}
where $2n^2>3m^2>0$. Time evolution of the spin breather is shown in  \fref{fig:spin_gab}, for $n=5$ and $m = 2$. Equally spaced excitations can be seen during the breather excitation which will eventually vanish as $t \rightarrow \pm\infty$. The localized excitations may be compared with that of the corresponding breather profile $\psi_{\rm GAB}$, shown in \fref{fig:psi_gab}. The uniform rotation in time present in the background spin field $\uvec{S}_{\rm h}$, \eref{Sh}, is cancelled out in the analysis for simplicity. This will not alter the conclusion to be drawn from the analysis. It is to be noted that, in order to emphasise the effect of breather excitation in the phase evolution of AB, the time dependent part ($e^{it}$) has been cancelled out in \,\cite{devine:2011}. We are studying the recurrence in classical HF spin chain to see how close the recurrence is. One can see that the initial and final states of the spin chain shown in \fref{fig:spin_gab} (a) and (i) are not identical. They differ by a rotation about the lattice axis. Therefore, the recurrence is not exact. 

\begin{figure}[ht]
	\begin{minipage}[b]{0.23\textwidth}
	   \centering
	   \input{shift_h.tex}
	   \footnotesize{(a) Phase shift in $\psi_{\rm GAB}$}
	\end{minipage}
 \hfill 	
	\begin{minipage}[b]{0.23\textwidth}
	   \centering
	  \input{shift_c.tex}
	  \footnotesize{(b) Phase shift in $\psi_{\rm AB}$}
	\end{minipage}
\caption[Phase shift $\Delta \varphi$ and Additional global rotation $\Delta \theta$ as a function of $\lambda_{0I}$ ]{Phase shift $\Delta \varphi$ is the additional phase developed in the complex field $\psi$ during a breathing event. Additional global rotation picked up by the spin chain is denoted as $\Delta \theta$. Phase shift $\Delta \varphi$ (and $\Delta \theta$) versus $\lambda_{0I}$ is drawn for fixed $\kappa_0 (=1)$. (a) Galilean transformed AB and its spin counter part shows a shift in terms of $\Delta \varphi$ and $\Delta \theta$ during a recurrence. (b) Even though AB shows a non zero phase shift, the corresponding spin chain does not show any additional global rotation during the recurrence. In both cases (a) and (b), limit $\lambda_{0I} \to \kappa_0$ leads to the rogue event marked by $2\pi$ phase shift in the complex field. 
Breather solution does not exist in the limit $\lambda_{0I} \to 0$.}
\label{fig:shift}
\end{figure}
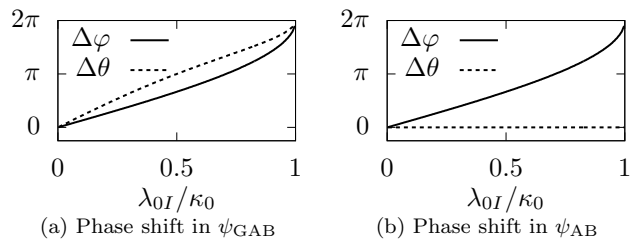

We denote the additional rotation picked up by the spin chain during the recurrence process as $\Delta \theta$. This is the difference in angle about the lattice axis, of any two spin vectors  at $(x_0, t_0)$ and $(x_0, -t_0)$, where the constant $t_0$ is relatively large so that $ t_0 \sim \pm\infty$. The complex breather solution reveals a definite shift it has undergone during a breathing event, via its phase factor, which may be written as 
\begin{equation}
\Delta\varphi = \mathrm{arg}(\psi(x_0, t_0))-\mathrm{arg}(\psi(x_0, -t_0)).
\end{equation}
Keeping $\kappa_0$ fixed, both the shifts $\Delta \theta$ and $\Delta \varphi$ are calculated as a function of $\lambda_{0I}$, and shown in \fref{fig:shift} as a comparison between two kinds of breathers, $\psi_{\rm GAB}$ and $\psi_{\rm AB}$ . In the limit, $\lambda_{0I} \to 0$, the breather excitation vanishes and hence no phase shift. On the other hand, $\lambda_{0I} \to \kappa_0$ leads to the rogue event indicated by $2\pi$ phase shift. Such a phase shift in the breather solution is expected\,\cite{devine:2011,kuznetsov:2017}. However, in view of the gauge equivalence between NLSE and the classical HF spin chain, a similar behaviour is expected in the spin chain. Surprisingly, the spin chain associated with  $\psi_{\rm AB}$ does not show any shift during the recurrence in contrast to that of $\psi_{\rm GAB}$. 

\begin{figure}[]
\centering
	\input{psi_gab.tex}
	\caption{Breather profile of Galilean transformed AB, \eref{psigab}. For comparison, the parameter values are same as that of the corresponding spin breather shown in \fref{fig:spin_gab}. Energy profile of the spin chain $E = \uvec{S}_x^2 = 4 |\psi|^2$ is qualitatively similar.}
\label{fig:psi_gab}
\end{figure}
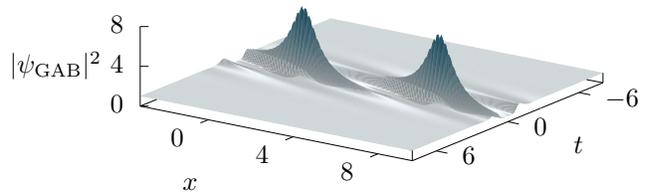

A dynamic field is said to exhibit recurrence when the initial state of the system is recovered in time either completely or as a close approximation. It can even be a repeated process. 
A recurrence process can be exactly modelled by Akhmediev breather (AB)\,\cite{akhm:2001}. In this framework a recurrence process receives certain unusual properties in contrast to a mere periodic process. One is that the recurrence takes place only once. Moreover, as $ t \rightarrow \pm\infty$, the system asymptotically attains (nearly) identical states; a kind of settled states free of any fluctuations. The ruffling occurs only for a short duration of time. However, for AB, there exist a measurable quantity, phase of the complex field, that distinguishes the initial and final state of the system\,\cite{devine:2011}. Therefore, some `trace' of the recurrence is left in the system in the form of phase shift.

\section{Conclusion}

We have shown in this work, two kinds of spatially periodic breathers in the HF spin chain: one associated with the AB, and the other associated with a Galilean transformed AB. In the former, the recurrence is exact, in the sense that the initial and final configuration are indistinguishable (\fref{fig:spin_ab}). In the later, the spin chain has undergone an additional global rotation, which clearly distinguish the initial and final states of the spin chain (\fref{fig:spin_gab}). 
It is worth mentioning that, both the breathers, AB as well as Galilean transformed AB, nevertheless show a phase shift during the recurrence process which makes the initial and final states distinguishable. 
Our result shows that a one-to-one correspondence between HF model and the NLSE, seems missing in view of the recurrence phenomena.
This issue has to be addressed either by suggesting a reason behind this ambiguity or by identifying the proper counterpart for the phase shift ($\Delta\varphi$) in HF spin chain.


\appendix
\section{Darboux transformation technique}
\label{app:DT}
The 1-d NLSE given in \eref{nlse}, arises as the compatibility condition for the linear system, also known as Lax pair given by \eref{laxp_nlse}, 
where the connections $U$ and $V$ are in general functions of $x,t$ and spectral parameter $\lambda$. For convenience one can treat them as functions of $\psi$ and its derivatives, as given below:

\begin{eqnarray}
U(\psi) = 
\bmat 	0 & \psi \\ -\overline{\psi} & 0 	\emat
+ \lambda 
\bmat  -i & 0 \\ 0 & i \emat , 
\nonumber \\
V(\psi) = 
\bmat  i{|\psi|}^2 & i \psi_x \\ i \overline{\psi}_x & -i{|\psi|}^2  \emat \nonumber \\
\qquad \quad
+ \lambda
\bmat 	0 & 2\psi \\ -2\overline{\psi} & 0 	\emat
+ \lambda^2
\bmat 	-2i & 0 \\ 0 & 2i 	\emat .
\end{eqnarray}
\subsection{Seed solution and seed spin}
\label{app:seed}
Seed solution $\psi_{\rm h} = \kappa_0 e^{i\sqrt{2}\, \kappa_0 x}$ corresponds to a matrix solution $\bm{\Psi}_{\rm h}(x,t,\lambda)$ to the below Lax pair
\begin{equation}
\label{appH:lph}
\eqalign{
\bm{\Psi}_{{\rm h},x} &= U_{\rm h} \bm{\Psi}_{\rm h} \ , \\
\bm{\Psi}_{{\rm h},t} &= V_{\rm h} \bm{\Psi}_{\rm h} \ ,
}
\end{equation}
where $U_{\rm h} = U(\psi_{\rm h})$, and $V_{\rm h}=V(\psi_{\rm h})$. 
We explicitly write the matrix solution $\bm{\Psi}_{\rm h}$ in the form,
\begin{equation}
\label{appH:Psih}
\bm{\Psi}_{\rm h}(x,t,\lambda) =   \frac{1}{\sqrt{d}}\	\bmat
						\varphi_1 & -\overline{\varphi}_2 \\ \varphi_2 \, & \, \, \overline{\varphi}_1
						\emat ,
\end{equation}
wherein, \\
$	\varphi_1 = \Big(a\, e^{i \Omega/2} + b\, e^{-i \Omega/2}  \Big) 
									e^{i \frac{1}{\sqrt{2}\,} \kappa_0 x} $ ,\\
$	\varphi_2 = - \Big(b\, e^{i \Omega/2} + a\, e^{-i \Omega/2}  \Big) 
									e^{-i \frac{1}{\sqrt{2}\,} \kappa_0 x} $ ,\\
$	\Omega 	= 2\, f \, ( x - \sqrt{2} \mu t ), \quad
	f 		= (\nicefrac{1}{\sqrt{2}}) \, \sqrt{\nu^2 + 2\, \kappa_0^2} $, \\
$	a = i ( \nu - \sqrt{2} f) - \sqrt{2} \kappa_0, \quad 
	b = i ( \nu + \sqrt{2} f) - \sqrt{2} \kappa_0 $	,	\\
$	d = 16 f^2 , \quad \mu = \kappa_0 - \sqrt{2}\, \lambda , \quad
	\nu = \kappa_0 + \sqrt{2}\, \lambda .$ \\
We have assumed the form of $\varphi_1$ and $\varphi_2$ in \eref{appH:Psih}, then substituted in \eref{appH:lph} to obtain its exact expression as given above.
The seed spin associated with $\psi_{\rm h}$, 
\begin{equation}
\label{Shmatrix}
\mathbf{S}_{\rm h} = \lim_{\lambda\to 0} {\bm{\Psi}_{\rm h}}^{\dagger} \sigma_3 \bm{\Psi}_{\rm h},
\end{equation}
is given in \eref{Sh} in the vector form.
\subsection{Darboux matrix $G_1(\lambda)$}
If $\upphi_{[1]}$ and $\upphi_{[2]}$ are two known vector valued eigenfunctions of the system \eref{appH:lph} corresponding to the parameters $\lambda_0$ and $\overline{\lambda}_0$. Using the matrices
$
M_0 = \mbox{diag}(\lambda_0, \overline{\lambda}_0), \ 
H = \big( \upphi_{[1]} \, \, \upphi_{[2]} \big), $ and $
G_0 = - H M_0 H^{-1},
$
the Darboux matrix $G_1(\lambda)$ can be written in the form $G_1(\lambda) = \lambda \mathbf{I} + G_0 $. 
We will first obtain suitable $\upphi_{[1]}$ and $\upphi_{[2]}$ from \eref{appH:Psih} by considering its column matrices as shown below:
\begin{equation} 
\label{appH:eigen}
\upphi_{[1]} = 	\frac{1}{\sqrt{d_0}}\ \bmat
				\upphi_1 \\ \upphi_2
				\emat , \ \  
\upphi_{[2]} =  \frac{1}{\sqrt{d_0}}\ \bmat
				-\overline{\upphi}_2 \\ \overline{\upphi}_1
				\emat ,
\end{equation}
wherein,\\
$ \upphi_1 	= \Big(a_0\, e^{i \Omega_0/2} + b_0\, e^{-i \Omega_0/2}  \Big) 
									e^{i \frac{1}{\sqrt{2}\,} \kappa_0 x}  	 $ ,\\
$ \upphi_2  = - \Big(b_0\, e^{i \Omega_0/2} + a_0\, e^{-i \Omega_0/2}  \Big) 
									e^{-i \frac{1}{\sqrt{2}\,} \kappa_0 x}	$ ,\\
$ \Omega_0 = 2\, f_0 \, ( x - \sqrt{2} \mu_0 t ) $ , 
$ f_0 		= (\nicefrac{1}{\sqrt{2}}) \, \sqrt{\nu_0^2 + 2\, \kappa_0^2} $ \\
$ a_0 = i ( \nu_0 - \sqrt{2} f_0) - \sqrt{2} \kappa_0 $ , 
$ b_0 = i ( \nu_0 + \sqrt{2} f_0) - \sqrt{2} \kappa_0 $ , 	\\
$ d_0 = 16 f_0^2, \ \mu_0 = \kappa_0 - \sqrt{2}\, \lambda_0 $ , 
$ \nu_0 = \kappa_0 + \sqrt{2}\, \lambda_0 . $ \\

One can verify that $\upphi_{[1]}$ and $\upphi_{[2]}$ defined in \eref{appH:eigen} satisfy 
\eref{appH:lph} corresponding to the parameters $\lambda_0$ and $\overline{\lambda}_0$ respectively.
A direct computation gives $G_0$ as,
\begin{eqnarray}
G_0 =   \bmat
		-\lambda_{0R} & 0 \\ 0 & -\lambda_{0R}
		\emat + 
		\frac{i \lambda_{0I}}{|\upphi_1|^2 +|\upphi_2|^2 } \nonumber \\
		\qquad \times 
		\bmat
		-(|\upphi_1|^2 -|\upphi_2|^2) & -2\upphi_1 \overline{\upphi}_2 \\
		-2\overline{\upphi}_1 \upphi_2  & (|\upphi_1|^2 -|\upphi_2|^2)
		\emat.
\end{eqnarray}
Simplification leads to:
\begin{eqnarray}
\label{appH:G0}
G_0 =   \bmat
		-\lambda_{0R} & 0 \\ 0 & -\lambda_{0R}
		\emat  
		+ \frac{i \lambda_{0I}}{\chi} 		\nonumber \\
		\quad \ \ \ \times
		\bmat
		-\xi & e^{i \sqrt{2} \kappa_0 x} (\zeta - i\, \eta ) \\
		 e^{- i \sqrt{2} \kappa_0 x} (\zeta + i\, \eta ) & \xi
		\emat,
\end{eqnarray}
where functions $\zeta, \eta, \xi$ and $\chi$ are defined below \eref{spin_hb}. 
As a remark --- $G_0$ is independent of $\lambda$, and $G_0 G_0^{\dagger} = |\lambda_0|^2 \bf{I}$.
The Darboux matrix $G_1(\lambda)$ is given by
\begin{equation}
G_1(\lambda) = \lambda \mathbf{I} + G_0 .
\end{equation}
\subsection{1-soliton (breather) solution}
\label{app:br}
Starting with a known solution $\psi_{\rm h}$, a new solution $\psi_{\rm hb}$ can be obtained by,
\begin{equation}
\psi_{\rm hb} = \psi_{\rm h} - 2i (G_0)_{12},
\end{equation}
where $(G_0)_{12}$ is the second element of the first row in $G_0$, \eref{appH:G0}. The breather solution  $\psi_{\rm hb}$ is given in \eref{psi_hb}. Note that $\psi_{\rm h}, -\psi_{\rm h}, \psi_{\rm hb}$ and $-\psi_{\rm hb}$ are all equally satisfy \eref{nlse}.
\subsection{Darboux transformation and spin breather}
\label{app:spinbr}
Darboux transformation gives a new $\bm{\Psi}_{\rm hb}$ by
\begin{equation}
\label{appH:psi1}
\bm{\Psi}_{\rm hb}(x,t,\lambda,\lambda_0) = \frac{1}{\sqrt{d_1}} G_1 (x,t,\lambda,\lambda_0) \, \bm{\Psi}_{\rm h} (x,t,\lambda) ,
\end{equation}
where $d_1 = |G_1|= (\lambda^2 + |\lambda_0|^2 - 2\lambda \lambda_{0R})$.
Matrix solution $\bm{\Psi}_{\rm hb}$ satisfies the Lax pair
\begin{equation}
\label{appH:lp1}
\eqalign{
\bm{\Psi}_{{\rm hb},x} &= U_{\rm hb} \bm{\Psi}_{\rm hb} \ , \\
\bm{\Psi}_{{\rm hb},t} &= V_{\rm hb} \bm{\Psi}_{\rm hb} \ ,
}
\end{equation}
where $U_{\rm hb} = U(\psi_{\rm hb})$, and $V_{\rm hb}=V(\psi_{\rm hb})$. One may compare \eref{appH:lp1}  with \eref{appH:lph} to see how Darboux transformation transform $\psi$ and $\bm{\Psi}$ systematically. Spin breather associated with $\bm{\Psi}_{\rm hb}$ is give by
\begin{equation}
\label{app:Shb}
\mathbf{S}_{\rm hb} = \lim_{\lambda\to 0} {\bm{\Psi}_{\rm hb}}^{\dagger} \sigma_3 \bm{\Psi}_{\rm hb},
\end{equation}
which is presented in \eref{spin_hb} in the vector form.

\section{Spatially periodic breather as a special case}
\label{app:GAB}
\noindent
We will show that the 1-breather 
 \eref{psi_hb} can be reduced to a spatially periodic breather \eref{psigab} using the condition  \eref{cond_GAB}.
Recall the functions defined below \eref{spin_hb}, and consider
\begin{equation}
2\, f_0^2 = \nu_0^2 + 2\, \kappa_0^2,
\label{fnot}
\end{equation}
where, $\nu_0 = \nu_{0R} + i\ \nu_{0I} = (\kappa_0 + \sqrt{2}\ \lambda_{0R}) + i \sqrt{2}\ \lambda_{0I}$.
Using \eref{cond_GAB} it follows that
\begin{equation}
\kappa_0^2 = \lambda_{0I}^2 + f_{0R}^2.
\label{fnotspecial}
\end{equation}
Introduce a real parameter $\phi$ such that,
\begin{equation}
\lambda_{0I} = \kappa_0 \cos \phi ,\ \mbox{and} \quad f_{0R} = \kappa_0  \sin \phi.
\end{equation}
This leads to $\mu_{0R} = 2\kappa_0$, $\mu_{0I} = -\sqrt{2}\lambda_{0I}$ and $f_{0I}=0$. It can be found that
\begin{equation}
\label{app:Omeg}
\Omega_{0R} = q(x-2\sqrt{2} \kappa_0 t), \quad
\Omega_{0I} = rt,
\end{equation}
where $q = 2 \kappa_0 \sin \phi $ and $ r = 2 \kappa_0^2 \sin(2\phi) $. Constants $c_i$ defined below \eref{spin_hb} now become,
$
c_1 = 16 (\kappa_0 + \lambda_{0I}) \kappa_0, \ c_2 = c_1 \cos \phi, \ c_3 = 0, $ and $ c_4 = -c_1 \sin \phi .
$
Functions $\zeta, \eta, \chi$  get simplified to
\begin{eqnarray}
\zeta &= c_1 \big( \cos(q(x-2\sqrt{2} \kappa_0 t)) + \cos \phi\ \cosh(rt) \big) , \nonumber \\
\eta &= c_1 \sin \phi\ \sinh(rt) ,\\
\chi &= c_1 \big( \cos \phi \ \cos(q(x-2\sqrt{2} \kappa_0 t)) + \cosh(rt) \big) . \nonumber
\end{eqnarray}
Making use of the substitutions:
\begin{eqnarray}
\cosh(rt-2i\phi) = \cosh(rt) \cos(2\phi) - i \sinh(rt) \sin(2\phi), \nonumber \\
v=2\sqrt{2} \kappa_0, \\
\cos(q(x-vt)) = -\cos(q(x-vt) + \pi), \nonumber 
\end{eqnarray}
and a coordinate shift
\begin{equation}
x \to x-x_0, \ \ \mbox{with} \ x_0=\pi/q,
\end{equation}
the breather $\psi_{\rm hb}$ \eref{psi_hb} can be reduced to $\psi_{\rm GAB}$ \eref{psigab}. 
It is clear that, $\psi_{\rm GAB}$ is localized in $t$ and periodic in $x$ due to \eref{app:Omeg}.

\section*{References}
\providecommand{\newblock}{}

\end{document}

%% file: psi_ab.tex
\begingroup
  \makeatletter
  \providecommand\color[2][]{%
    \GenericError{(gnuplot) \space\space\space\@spaces}{%
      Package color not loaded in conjunction with
      terminal option `colourtext'%
    }{See the gnuplot documentation for explanation.%
    }{Either use 'blacktext' in gnuplot or load the package
      color.sty in LaTeX.}%
    \renewcommand\color[2][]{}%
  }%
  \providecommand\includegraphics[2][]{%
    \GenericError{(gnuplot) \space\space\space\@spaces}{%
      Package graphicx or graphics not loaded%
    }{See the gnuplot documentation for explanation.%
    }{The gnuplot epslatex terminal needs graphicx.sty or graphics.sty.}%
    \renewcommand\includegraphics[2][]{}%
  }%
  \providecommand\rotatebox[2]{#2}%
  \@ifundefined{ifGPcolor}{%
    \newif\ifGPcolor
    \GPcolortrue
  }{}%
  \@ifundefined{ifGPblacktext}{%
    \newif\ifGPblacktext
    \GPblacktexttrue
  }{}%
  \let\gplgaddtomacro\g@addto@macro
  \gdef\gplbacktext{}%
  \gdef\gplfronttext{}%
  \makeatother
  \ifGPblacktext
    \def\colorrgb#1{}%
    \def\colorgray#1{}%
  \else
    \ifGPcolor
      \def\colorrgb#1{\color[rgb]{#1}}%
      \def\colorgray#1{\color[gray]{#1}}%
      \expandafter\def\csname LTw\endcsname{\color{white}}%
      \expandafter\def\csname LTb\endcsname{\color{black}}%
      \expandafter\def\csname LTa\endcsname{\color{black}}%
      \expandafter\def\csname LT0\endcsname{\color[rgb]{1,0,0}}%
      \expandafter\def\csname LT1\endcsname{\color[rgb]{0,1,0}}%
      \expandafter\def\csname LT2\endcsname{\color[rgb]{0,0,1}}%
      \expandafter\def\csname LT3\endcsname{\color[rgb]{1,0,1}}%
      \expandafter\def\csname LT4\endcsname{\color[rgb]{0,1,1}}%
      \expandafter\def\csname LT5\endcsname{\color[rgb]{1,1,0}}%
      \expandafter\def\csname LT6\endcsname{\color[rgb]{0,0,0}}%
      \expandafter\def\csname LT7\endcsname{\color[rgb]{1,0.3,0}}%
      \expandafter\def\csname LT8\endcsname{\color[rgb]{0.5,0.5,0.5}}%
    \else
      \def\colorrgb#1{\color{black}}%
      \def\colorgray#1{\color[gray]{#1}}%
      \expandafter\def\csname LTw\endcsname{\color{white}}%
      \expandafter\def\csname LTb\endcsname{\color{black}}%
      \expandafter\def\csname LTa\endcsname{\color{black}}%
      \expandafter\def\csname LT0\endcsname{\color{black}}%
      \expandafter\def\csname LT1\endcsname{\color{black}}%
      \expandafter\def\csname LT2\endcsname{\color{black}}%
      \expandafter\def\csname LT3\endcsname{\color{black}}%
      \expandafter\def\csname LT4\endcsname{\color{black}}%
      \expandafter\def\csname LT5\endcsname{\color{black}}%
      \expandafter\def\csname LT6\endcsname{\color{black}}%
      \expandafter\def\csname LT7\endcsname{\color{black}}%
      \expandafter\def\csname LT8\endcsname{\color{black}}%
    \fi
  \fi
    \setlength{\unitlength}{0.0500bp}%
    \ifx\gptboxheight\undefined%
      \newlength{\gptboxheight}%
      \newlength{\gptboxwidth}%
      \newsavebox{\gptboxtext}%
    \fi%
    \setlength{\fboxrule}{0.5pt}%
    \setlength{\fboxsep}{1pt}%
\begin{picture}(4176.00,1440.00)%
    \gplgaddtomacro\gplbacktext{%
    }%
    \gplgaddtomacro\gplfronttext{%
      \csname LTb\endcsname
      \put(966,300){\makebox(0,0)[r]{\strut{}$0$}}%
      \put(1489,196){\makebox(0,0)[r]{\strut{}$4$}}%
      \put(2011,91){\makebox(0,0)[r]{\strut{}$8$}}%
      \put(1012,-26){\makebox(0,0){\strut{}$x$}}%
      \put(4141,603){\makebox(0,0)[l]{\strut{}$-6$}}%
      \put(3602,383){\makebox(0,0)[l]{\strut{}$0$}}%
      \put(3063,163){\makebox(0,0)[l]{\strut{}$6$}}%
      \put(3939,251){\makebox(0,0){\strut{}$t$}}%
      \put(510,555){\makebox(0,0)[r]{\strut{}$0$}}%
      \put(510,855){\makebox(0,0)[r]{\strut{}$4$}}%
      \put(510,1155){\makebox(0,0)[r]{\strut{}$8$}}%
      \put(2,855){\makebox(0,0){\strut{}$|\psi_{\rm AB}|^2$}}%
    }%
    \gplbacktext
    \put(0,0){\includegraphics{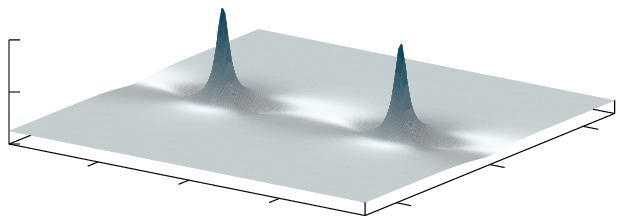}}%
    \gplfronttext
  \end{picture}%
\endgroup

%% file: psi_gab.tex
\begingroup
  \makeatletter
  \providecommand\color[2][]{%
    \GenericError{(gnuplot) \space\space\space\@spaces}{%
      Package color not loaded in conjunction with
      terminal option `colourtext'%
    }{See the gnuplot documentation for explanation.%
    }{Either use 'blacktext' in gnuplot or load the package
      color.sty in LaTeX.}%
    \renewcommand\color[2][]{}%
  }%
  \providecommand\includegraphics[2][]{%
    \GenericError{(gnuplot) \space\space\space\@spaces}{%
      Package graphicx or graphics not loaded%
    }{See the gnuplot documentation for explanation.%
    }{The gnuplot epslatex terminal needs graphicx.sty or graphics.sty.}%
    \renewcommand\includegraphics[2][]{}%
  }%
  \providecommand\rotatebox[2]{#2}%
  \@ifundefined{ifGPcolor}{%
    \newif\ifGPcolor
    \GPcolortrue
  }{}%
  \@ifundefined{ifGPblacktext}{%
    \newif\ifGPblacktext
    \GPblacktexttrue
  }{}%
  \let\gplgaddtomacro\g@addto@macro
  \gdef\gplbacktext{}%
  \gdef\gplfronttext{}%
  \makeatother
  \ifGPblacktext
    \def\colorrgb#1{}%
    \def\colorgray#1{}%
  \else
    \ifGPcolor
      \def\colorrgb#1{\color[rgb]{#1}}%
      \def\colorgray#1{\color[gray]{#1}}%
      \expandafter\def\csname LTw\endcsname{\color{white}}%
      \expandafter\def\csname LTb\endcsname{\color{black}}%
      \expandafter\def\csname LTa\endcsname{\color{black}}%
      \expandafter\def\csname LT0\endcsname{\color[rgb]{1,0,0}}%
      \expandafter\def\csname LT1\endcsname{\color[rgb]{0,1,0}}%
      \expandafter\def\csname LT2\endcsname{\color[rgb]{0,0,1}}%
      \expandafter\def\csname LT3\endcsname{\color[rgb]{1,0,1}}%
      \expandafter\def\csname LT4\endcsname{\color[rgb]{0,1,1}}%
      \expandafter\def\csname LT5\endcsname{\color[rgb]{1,1,0}}%
      \expandafter\def\csname LT6\endcsname{\color[rgb]{0,0,0}}%
      \expandafter\def\csname LT7\endcsname{\color[rgb]{1,0.3,0}}%
      \expandafter\def\csname LT8\endcsname{\color[rgb]{0.5,0.5,0.5}}%
    \else
      \def\colorrgb#1{\color{black}}%
      \def\colorgray#1{\color[gray]{#1}}%
      \expandafter\def\csname LTw\endcsname{\color{white}}%
      \expandafter\def\csname LTb\endcsname{\color{black}}%
      \expandafter\def\csname LTa\endcsname{\color{black}}%
      \expandafter\def\csname LT0\endcsname{\color{black}}%
      \expandafter\def\csname LT1\endcsname{\color{black}}%
      \expandafter\def\csname LT2\endcsname{\color{black}}%
      \expandafter\def\csname LT3\endcsname{\color{black}}%
      \expandafter\def\csname LT4\endcsname{\color{black}}%
      \expandafter\def\csname LT5\endcsname{\color{black}}%
      \expandafter\def\csname LT6\endcsname{\color{black}}%
      \expandafter\def\csname LT7\endcsname{\color{black}}%
      \expandafter\def\csname LT8\endcsname{\color{black}}%
    \fi
  \fi
    \setlength{\unitlength}{0.0500bp}%
    \ifx\gptboxheight\undefined%
      \newlength{\gptboxheight}%
      \newlength{\gptboxwidth}%
      \newsavebox{\gptboxtext}%
    \fi%
    \setlength{\fboxrule}{0.5pt}%
    \setlength{\fboxsep}{1pt}%
\begin{picture}(4176.00,1440.00)%
    \gplgaddtomacro\gplbacktext{%
    }%
    \gplgaddtomacro\gplfronttext{%
      \csname LTb\endcsname
      \put(966,300){\makebox(0,0)[r]{\strut{}$0$}}%
      \put(1606,172){\makebox(0,0)[r]{\strut{}$4$}}%
      \put(2246,44){\makebox(0,0)[r]{\strut{}$8$}}%
      \put(1012,-26){\makebox(0,0){\strut{}$x$}}%
      \put(4141,603){\makebox(0,0)[l]{\strut{}$-6$}}%
      \put(3602,383){\makebox(0,0)[l]{\strut{}$0$}}%
      \put(3063,163){\makebox(0,0)[l]{\strut{}$6$}}%
      \put(3939,251){\makebox(0,0){\strut{}$t$}}%
      \put(510,555){\makebox(0,0)[r]{\strut{}$0$}}%
      \put(510,855){\makebox(0,0)[r]{\strut{}$4$}}%
      \put(510,1155){\makebox(0,0)[r]{\strut{}$8$}}%
      \put(2,855){\makebox(0,0){\strut{}$|\psi_{\rm GAB}|^2$}}%
    }%
    \gplbacktext
    \put(0,0){\includegraphics{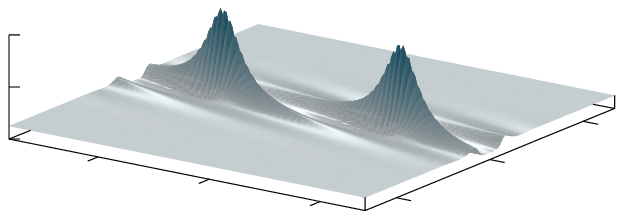}}%
    \gplfronttext
  \end{picture}%
\endgroup

%% file: shift_h.tex
\begingroup
  \makeatletter
  \providecommand\color[2][]{%
    \GenericError{(gnuplot) \space\space\space\@spaces}{%
      Package color not loaded in conjunction with
      terminal option `colourtext'%
    }{See the gnuplot documentation for explanation.%
    }{Either use 'blacktext' in gnuplot or load the package
      color.sty in LaTeX.}%
    \renewcommand\color[2][]{}%
  }%
  \providecommand\includegraphics[2][]{%
    \GenericError{(gnuplot) \space\space\space\@spaces}{%
      Package graphicx or graphics not loaded%
    }{See the gnuplot documentation for explanation.%
    }{The gnuplot epslatex terminal needs graphicx.sty or graphics.sty.}%
    \renewcommand\includegraphics[2][]{}%
  }%
  \providecommand\rotatebox[2]{#2}%
  \@ifundefined{ifGPcolor}{%
    \newif\ifGPcolor
    \GPcolortrue
  }{}%
  \@ifundefined{ifGPblacktext}{%
    \newif\ifGPblacktext
    \GPblacktexttrue
  }{}%
  \let\gplgaddtomacro\g@addto@macro
  \gdef\gplbacktext{}%
  \gdef\gplfronttext{}%
  \makeatother
  \ifGPblacktext
    \def\colorrgb#1{}%
    \def\colorgray#1{}%
  \else
    \ifGPcolor
      \def\colorrgb#1{\color[rgb]{#1}}%
      \def\colorgray#1{\color[gray]{#1}}%
      \expandafter\def\csname LTw\endcsname{\color{white}}%
      \expandafter\def\csname LTb\endcsname{\color{black}}%
      \expandafter\def\csname LTa\endcsname{\color{black}}%
      \expandafter\def\csname LT0\endcsname{\color[rgb]{1,0,0}}%
      \expandafter\def\csname LT1\endcsname{\color[rgb]{0,1,0}}%
      \expandafter\def\csname LT2\endcsname{\color[rgb]{0,0,1}}%
      \expandafter\def\csname LT3\endcsname{\color[rgb]{1,0,1}}%
      \expandafter\def\csname LT4\endcsname{\color[rgb]{0,1,1}}%
      \expandafter\def\csname LT5\endcsname{\color[rgb]{1,1,0}}%
      \expandafter\def\csname LT6\endcsname{\color[rgb]{0,0,0}}%
      \expandafter\def\csname LT7\endcsname{\color[rgb]{1,0.3,0}}%
      \expandafter\def\csname LT8\endcsname{\color[rgb]{0.5,0.5,0.5}}%
    \else
      \def\colorrgb#1{\color{black}}%
      \def\colorgray#1{\color[gray]{#1}}%
      \expandafter\def\csname LTw\endcsname{\color{white}}%
      \expandafter\def\csname LTb\endcsname{\color{black}}%
      \expandafter\def\csname LTa\endcsname{\color{black}}%
      \expandafter\def\csname LT0\endcsname{\color{black}}%
      \expandafter\def\csname LT1\endcsname{\color{black}}%
      \expandafter\def\csname LT2\endcsname{\color{black}}%
      \expandafter\def\csname LT3\endcsname{\color{black}}%
      \expandafter\def\csname LT4\endcsname{\color{black}}%
      \expandafter\def\csname LT5\endcsname{\color{black}}%
      \expandafter\def\csname LT6\endcsname{\color{black}}%
      \expandafter\def\csname LT7\endcsname{\color{black}}%
      \expandafter\def\csname LT8\endcsname{\color{black}}%
    \fi
  \fi
    \setlength{\unitlength}{0.0500bp}%
    \ifx\gptboxheight\undefined%
      \newlength{\gptboxheight}%
      \newlength{\gptboxwidth}%
      \newsavebox{\gptboxtext}%
    \fi%
    \setlength{\fboxrule}{0.5pt}%
    \setlength{\fboxsep}{1pt}%
\begin{picture}(2160.00,1440.00)%
    \gplgaddtomacro\gplbacktext{%
      \csname LTb\endcsname
      \put(192,604){\makebox(0,0)[r]{\strut{}$0$}}%
      \put(192,1007){\makebox(0,0)[r]{\strut{}$\pi$}}%
      \put(192,1410){\makebox(0,0)[r]{\strut{}$2\pi$}}%
      \put(324,283){\makebox(0,0){\strut{}$0$}}%
      \put(1220,283){\makebox(0,0){\strut{}$0.5$}}%
      \put(2115,283){\makebox(0,0){\strut{}$1$}}%
    }%
    \gplgaddtomacro\gplfronttext{%
      \csname LTb\endcsname
      \put(1219,63){\makebox(0,0){\strut{}$\lambda_{0I}/\kappa_0$}}%
      \csname LTb\endcsname
      \put(721,1261){\makebox(0,0)[r]{\strut{}$\Delta \varphi $}}%
      \csname LTb\endcsname
      \put(721,1036){\makebox(0,0)[r]{\strut{}$\Delta \theta $}}%
    }%
    \gplbacktext
    \put(0,0){\includegraphics{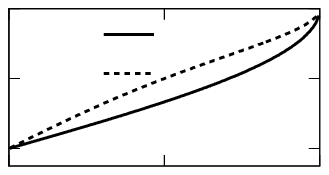}}%
    \gplfronttext
  \end{picture}%
\endgroup

%% file: shift_c.tex
\begingroup
  \makeatletter
  \providecommand\color[2][]{%
    \GenericError{(gnuplot) \space\space\space\@spaces}{%
      Package color not loaded in conjunction with
      terminal option `colourtext'%
    }{See the gnuplot documentation for explanation.%
    }{Either use 'blacktext' in gnuplot or load the package
      color.sty in LaTeX.}%
    \renewcommand\color[2][]{}%
  }%
  \providecommand\includegraphics[2][]{%
    \GenericError{(gnuplot) \space\space\space\@spaces}{%
      Package graphicx or graphics not loaded%
    }{See the gnuplot documentation for explanation.%
    }{The gnuplot epslatex terminal needs graphicx.sty or graphics.sty.}%
    \renewcommand\includegraphics[2][]{}%
  }%
  \providecommand\rotatebox[2]{#2}%
  \@ifundefined{ifGPcolor}{%
    \newif\ifGPcolor
    \GPcolortrue
  }{}%
  \@ifundefined{ifGPblacktext}{%
    \newif\ifGPblacktext
    \GPblacktexttrue
  }{}%
  \let\gplgaddtomacro\g@addto@macro
  \gdef\gplbacktext{}%
  \gdef\gplfronttext{}%
  \makeatother
  \ifGPblacktext
    \def\colorrgb#1{}%
    \def\colorgray#1{}%
  \else
    \ifGPcolor
      \def\colorrgb#1{\color[rgb]{#1}}%
      \def\colorgray#1{\color[gray]{#1}}%
      \expandafter\def\csname LTw\endcsname{\color{white}}%
      \expandafter\def\csname LTb\endcsname{\color{black}}%
      \expandafter\def\csname LTa\endcsname{\color{black}}%
      \expandafter\def\csname LT0\endcsname{\color[rgb]{1,0,0}}%
      \expandafter\def\csname LT1\endcsname{\color[rgb]{0,1,0}}%
      \expandafter\def\csname LT2\endcsname{\color[rgb]{0,0,1}}%
      \expandafter\def\csname LT3\endcsname{\color[rgb]{1,0,1}}%
      \expandafter\def\csname LT4\endcsname{\color[rgb]{0,1,1}}%
      \expandafter\def\csname LT5\endcsname{\color[rgb]{1,1,0}}%
      \expandafter\def\csname LT6\endcsname{\color[rgb]{0,0,0}}%
      \expandafter\def\csname LT7\endcsname{\color[rgb]{1,0.3,0}}%
      \expandafter\def\csname LT8\endcsname{\color[rgb]{0.5,0.5,0.5}}%
    \else
      \def\colorrgb#1{\color{black}}%
      \def\colorgray#1{\color[gray]{#1}}%
      \expandafter\def\csname LTw\endcsname{\color{white}}%
      \expandafter\def\csname LTb\endcsname{\color{black}}%
      \expandafter\def\csname LTa\endcsname{\color{black}}%
      \expandafter\def\csname LT0\endcsname{\color{black}}%
      \expandafter\def\csname LT1\endcsname{\color{black}}%
      \expandafter\def\csname LT2\endcsname{\color{black}}%
      \expandafter\def\csname LT3\endcsname{\color{black}}%
      \expandafter\def\csname LT4\endcsname{\color{black}}%
      \expandafter\def\csname LT5\endcsname{\color{black}}%
      \expandafter\def\csname LT6\endcsname{\color{black}}%
      \expandafter\def\csname LT7\endcsname{\color{black}}%
      \expandafter\def\csname LT8\endcsname{\color{black}}%
    \fi
  \fi
    \setlength{\unitlength}{0.0500bp}%
    \ifx\gptboxheight\undefined%
      \newlength{\gptboxheight}%
      \newlength{\gptboxwidth}%
      \newsavebox{\gptboxtext}%
    \fi%
    \setlength{\fboxrule}{0.5pt}%
    \setlength{\fboxsep}{1pt}%
\begin{picture}(2160.00,1440.00)%
    \gplgaddtomacro\gplbacktext{%
      \csname LTb\endcsname
      \put(192,604){\makebox(0,0)[r]{\strut{}$0$}}%
      \put(192,1007){\makebox(0,0)[r]{\strut{}$\pi$}}%
      \put(192,1410){\makebox(0,0)[r]{\strut{}$2\pi$}}%
      \put(324,283){\makebox(0,0){\strut{}$0$}}%
      \put(1220,283){\makebox(0,0){\strut{}$0.5$}}%
      \put(2115,283){\makebox(0,0){\strut{}$1$}}%
    }%
    \gplgaddtomacro\gplfronttext{%
      \csname LTb\endcsname
      \put(1219,63){\makebox(0,0){\strut{}$\lambda_{0I}/\kappa_0$}}%
      \csname LTb\endcsname
      \put(721,1261){\makebox(0,0)[r]{\strut{}$\Delta \varphi $}}%
      \csname LTb\endcsname
      \put(721,1036){\makebox(0,0)[r]{\strut{}$\Delta \theta $}}%
    }%
    \gplbacktext
    \put(0,0){\includegraphics{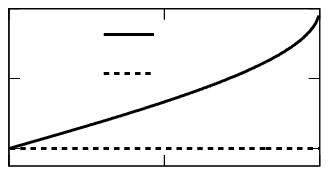}}%
    \gplfronttext
  \end{picture}%
\endgroup